\documentclass[aps, prd, twocolumn, preprintnumbers, nofootinbib]{revtex4}

\usepackage{amsmath}
\usepackage{bm}
\usepackage{graphicx}
\usepackage{epstopdf}
\usepackage{hyperref}
\usepackage{soul}
\usepackage{color}

\enlargethispage{\baselineskip}

\hypersetup{
     colorlinks   = true,
     citecolor    = blue
}

\newcommand{\MPl}{M_{\rm Pl}}
\newcommand{\Lb}{\Lambda_{\rm bulk}}

\newcommand{\dtm}{\delta t_{\rm meas}}

\newcommand{\be}{\begin{equation}}
\newcommand{\ee}{\end{equation}}
\newcommand{\bea}{\begin{eqnarray}}
\newcommand{\eea}{\end{eqnarray}}
\renewcommand\({\left(}
\renewcommand\){\right)}
\renewcommand\[{\left[}
\renewcommand\]{\right]}

\begin{document}

\newcommand{\FIRSTAFF}{\affiliation{The Oskar Klein Centre for Cosmoparticle Physics, Department of Physics, Stockholm University, \\AlbaNova Universitetscentrum, Roslagstullbacken 21A, SE-106 91 Stockholm, Sweden}}
\newcommand{\SECONDAFF}{\affiliation{Nordita, KTH Royal Institute of Technology and Stockholm University, Roslagstullsbacken 23, SE-106 91 Stockholm, Sweden}}
\newcommand{\THIRDAFF}{\affiliation{Central European Institute for Cosmology and Fundamental Physics (CEICO), \\Fyzik\'{a}ln\'{i} \'{u}stav Akademie v\v{e}d \v{C}R, Na Slovance 1999/2, CZ-182 21 Praha 8, Czech Republic}}

\title{Brane-world extra dimensions in light of GW170817}

\author{Luca Visinelli}
\email[Electronic address: ]{luca.visinelli@fysik.su.se}
\FIRSTAFF
\SECONDAFF

\author{Nadia Bolis}
\email[Electronic address: ]{bolis@fzu.cz, nbolis@ucdavis.edu}
\THIRDAFF

\author{Sunny Vagnozzi}
\email[Electronic address: ]{sunny.vagnozzi@fysik.su.se}
\FIRSTAFF

\date{\today}

\begin{abstract}
The search for extra dimensions is a challenging endeavor to probe physics beyond the Standard Model. The joint detection of gravitational waves (GW) and electromagnetic (EM) signals from the merging of a binary system of compact objects like neutron stars (NS), can help constrain the geometry of extra dimensions beyond our 3+1 spacetime ones. A theoretically well-motivated possibility is that our observable Universe is a 3+1-dimensional hypersurface, or brane, embedded in a higher 4+1-dimensional Anti-de Sitter (AdS$_5$) spacetime, in which gravity is the only force which propagates through the infinite bulk space, while other forces are confined to the brane. In these types of brane-world models, GW and EM signals between two points on the brane would, in general, travel different paths. This would result in a time-lag between the detection of GW and EM signals emitted simultaneously from the same source. We consider the recent near-simultaneous detection of the GW event GW170817 from the LIGO/Virgo collaboration, and its EM counterpart, the short gamma-ray burst GRB170817A detected by the \textit{Fermi} Gamma-ray Burst Monitor and the \textit{INTEGRAL} Anti-Coincidence Shield spectrometer. Assuming the standard $\Lambda$-Cold Dark Matter ($\Lambda$CDM) scenario and performing a likelihood analysis which takes into account astrophysical uncertainties associated to the measured time-lag, we set an upper limit of $\ell \lesssim 0.535\,$Mpc at $68\%$ confidence level on the AdS$_5$ radius of curvature $\ell$. Although the bound is not competitive with current Solar System constraints, it is the first time that data from a multi-messenger GW-EM measurement is used to constrain extra-dimensional models. Thus, our work provides a proof-of-principle for the possibility of using multi-messenger astronomy for probing the geometry of our space-time.
\end{abstract}
\maketitle

\section{Introduction}

The era of Gravitational Wave (GW) astronomy has come to its mature stage, following the first detections of GWs from binary black hole (BH) mergers~\cite{merger1, merger2, merger3, merger4, merger5} by the Advanced Laser Interferometer Gravitational Wave Observatory (LIGO)~\cite{aasi2015} and, more recently, by the Advanced Virgo~\cite{acernese2015} detectors. These discoveries were recently awarded the 2017 Nobel Prize in physics and sparked the search for exciting astrophysical phenomena~\cite{hartwig2016, giudice2016, maselli2016, sousa2016, addazi2017}. The use of three interferometers dramatically improves the localization of the source, as the detection of the event GW170814 showed~\cite{merger5}.

Besides BHs, stellar compact objects such as neutron stars (NS) can also be seen through the detection of GWs released when NS-NS or NS-BH binaries merge. Compact object binaries have since long been of interest as they provide important constraints on the structure and the formation of these extreme objects. Among the first NS binaries observed, one included a radio pulsar partner~\cite{hulse1975} whose close orbit was later observed to decay as predicted by Einstein's General Relativity. This observation was an indirect confirmation of the existence of GWs and was awarded the 1993 Nobel Prize in physics. The collision of NS binaries might also explain the short-duration gamma-ray bursts (SGRBs) observed~\cite{piran2004, kumar2015}.

The direct detection of GWs from a NS-NS merger, the event GW170817 in NGC 4993, has just been announced by the LIGO-Virgo network~\cite{NS2017, NS20171} with very high significance. The masses of the involved NSs have been measured to be respectively $M_1 = (1.36-1.60)$ and $M_2 = (1.17-1.36)$ solar masses. The \textit{Fermi} Gamma-ray Burst Monitor and the Anti-Coincidence Shield of the the International Gamma-Ray Astrophysics Laboratory (\textit{INTEGRAL}) spectrometer have also just detected a SGRB transient in NGC 4993, the event GRB170817A~\cite{NS2017, NS20171, Goldstein2017}, which has been associated to the GW event from the NS-NS merger with high significance. The associated energetics, variability and light curves, have also been shown to be highly consistent with a NS-NS merger~\cite{berthier2017}. The transient signal in the optical has been independently observed within an hour from the detection of the GW and SGRB events by the SWOPE telescope, allowing for a precise location of the host galaxy~\cite{NS2017, Kilpatrick:2017}. Further observations later detected the signal in the X-ray and radio wave spectra. The time delay between the detection of the GW and the SGRB counterpart shows a measured time-lag $\dtm = (1.734 \pm 0.054)\,$s, with the observed duration of the gamma-ray emission lasting $(2.0 \pm 0.5)\,$s, consistent with what is expected from a SGRB.

The combination of GW and EM signals from a binary merger can be used to probe the geometry of extra dimensions beyond our 3+1 spacetime ones. The possibility that additional dimensions exist was first postulated by Kaluza and Klein~\cite{kaluza1921,klein1926a, klein1926b} while attempting to unify gravity and electromagnetism. An intriguing possibility is that our observable spacetime is actually a 3+1 hypersurface (or {\it brane}) embedded in a higher 4+1 dimensional bulk space. The idea that our 3+1 spacetime is a boundary or brane of a higher dimensional space was brought up in the Ho\v{r}ava-Witten theory~\cite{horava1996}, and was soon used as an attempt to explain the mass scale hierarchy problem~\cite{arkanihamed1998}. Randall and Sundrum~\cite{randall1999} proposed a five-dimension AdS space (AdS$_5$) brane-world model where the extra dimension has an infinite size and a negative bulk cosmological constant $\Lb$. The tension on the brane $\sigma$ is tuned so that general relativity is recovered in the low-energy regime~\cite{binetruy1, binetruy2}. Such extra dimensions, not necessarily warped, can also yield important astrophysical consequences~\cite{arkanihamed1999, antoniadis1999, barger1999, hall1999}. See e.g.~\cite{maartens2010} for a review on brane-world gravity models.

In a subset of theories with extra dimensions, gravity propagates through the bulk while other fields like matter or radiation are confined to travel on the brane. In a seminal paper, Chung and Freese~\cite{chung1999} showed that gravitational waves might travel along an extra-dimensional null geodesic, so that a GW signal would reach us in a shorter time compared to a luminous signal emitted simultaneously, the latter being constrained to propagate on our three-brane. In short, null-geodesics in the five-dimensional space might causally connect two spacetime points A and B otherwise not in causal contact on the usual four-dimensional manifold. As a result, there might be a lag in the measurement of GW and EM signals emitted from the same source. A measurement of differing speeds of GW and EM signals might then arise from the existence of an extra dimension other than our 3+1 spacetime ones.


Causality within the brane Universe has been further discussed in Refs.~\cite{caldwell2001, ishihara2001, abdalla2001}, while additional work on GW in extra-dimensions was explored in Refs.~\cite{clarkson2007, garca2014, garciabellido2016, sumanta}. Solar System tests show that Newtonian gravity is in agreement with data down to scales of the order of a tenth of a millimeter~\cite{long2003, iorio2012, linares2014, tan2016}.

In this article we follow closely the calculation done by Caldwell and Langlois in Ref.~\cite{caldwell2001}, restricting ourselves to the case of a $\Lambda$-Cold Dark Matter ($\Lambda$CDM) scenario. Using the time delay between the signals detected by LIGO/Virgo and \textit{Fermi}/\textit{INTEGRAL}, we set a limit on the radius of curvature of the extra dimension.

This paper is structured as follows: in Sec.~(\ref{sec:calculations}) we compute the distance travelled by a GW signal along an extra dimensional geodesic, and hence the expected time-lag between the GW and the corresponding EM signal. In Sec.~(\ref{sec:analysis}), we describe the analysis method we use to analyse the time-lag measured in conjunction with the GW170817 event and constrain the physics of extra dimensions. In Sec.~(\ref{sec:results}) we present our results, and in particular the bounds on the AdS$_5$ radius of curvature $\ell$. Conclusions are drawn in Sec.~(\ref{sec:conclusions}).

\section{Computing the shortcut length}
\label{sec:calculations}

Following the set up in Ref.~\cite{caldwell2001} we consider a space-time metric analogous to the Friedmann-Robertson-Walker (FRW) metric, which describes the evolution of the Universe as a three-brane embedded in a five-dimensional, infinite, Anti-deSitter (AdS$_5$) space~\cite{binetruy1, binetruy2}. We assume that I) the three-brane representing our Universe is homogeneous and isotropic, and II) the branes are sufficiently distant so that we can approximate the bulk as being empty. Under these assumptions, the metric is given by
\be
	ds^2 = -f(R) dT^2 + f^{-1}(R)dR^2 + R^2d\Sigma^2,
	\label{eq:RSmetric}
\ee
where $d\Sigma^2$ is the maximally-symmetric metric for a three-dimensional space and $f(R)$ is a scale factor. Following the language of general relativity, the coordinates ($T$, $R$) are referred to as ``curvature coordinates'', with the time-like coordinate $T$ being the ``Killing time''~\cite{brill1997}.

Defining the proper time for the comoving observers on the three-brane
\be
	dt^2 = f(R)dT^2 - \frac{dR^2}{f(R)},
	\label{eq:propertime}
\ee
the metric describing the motion of a particle confined to the three-brane is $ds_{\rm brane}^2 = -dt^2 + R^2d\Sigma^2$, corresponding to a FRW metric in four dimensions with the scale factor $R(T) \equiv a(t)$. In this model, the expansion rate differs from the usual Friedmann expression~\cite{binetruy1, binetruy2}, which is recovered in the low energy limit $\rho \ll \sigma$, where $\rho$ is the matter energy density and $\sigma$ is the brane tension, as
\be
	\(\frac{\dot a}{a}\)^2 = \frac{\rho}{3\MPl^2}, \quad \hbox{(low-energy limit)}.
    \label{eq:hubblerate_std}
\ee
We have introduced the reduced Planck mass $\MPl = 1/\sqrt{8\pi G}$.

We assume that a GW signal is emitted at some point A on the brane, travels on a radial null-geodesic through the bulk, and is received at point B also lying on the brane. Defining the Hubble rate $H = (dR/dt)/R$, the Killing time spent when traveling between the source and the detector is
\be
	T_B - T_A = \int_{t_A}^{t_B} dt' \sqrt{\frac{1}{f(R)} + \frac{H^2R^2}{f^2(R)}}.
	\label{eq:timesplit}
\ee
We assume a flat three-dimensional space for which $d\Sigma^2 = dr^2+r^2d\Omega^2$, where $r$ is the radial coordinate and $d\Omega$ is the differential solid angle.  The scale factor is $f(R) = (R/\ell)^2$, and $\ell$ defines the constant AdS$_5$ curvature radius.~\footnote{An additional term $\mu/R^2$ can be added to the scale factor $f(R)$, with the term $\mu$ describing the mass of a black hole for a Schwarzschild-like solution of the AdS$_5$ metric~\cite{witten1998}. For simplicity we have neglected the possibility that a black hole affects the metric in the bulk by setting $\mu = 0$ throughout the present work.} We focus on radial geodesics, for which the metric in Eq.~\eqref{eq:RSmetric} reduces to 
\be
	ds^2 = -f(R) dT^2 + f^{-1}(R)dR^2 + R^2dr^2.
	\label{eq:radial_geodesic}
\ee
Given an affine parameter $\lambda$, this geodesic allows for two Killing vectors, with conserved quantities $E$ and $P$,
\be
	k_T = - f(R)\frac{dT}{d\lambda} = -E,\quad \hbox{and} \quad k_r = R^2\frac{dr}{d\lambda} = P.
\ee
For null geodesics $ds^2 = 0$, Eq.~\eqref{eq:radial_geodesic} with the conserved quantities above gives
\be
	\(\frac{dR}{d\lambda}\)^2 = E^2 - \frac{f(R)}{R^2}P^2
	\label{eq:nullgeodesic}
\ee
Combining Eq.~\eqref{eq:nullgeodesic} with the Killing vector $k_r$ gives us a relation between the distances on the three-brane and the radial coordinate in the five-dimensional space, which can be integrated from $R_A$ to $R_B$ to obtain ($r = r_B - r_A$)
\be
	\frac{1}{R_A} - \frac{1}{R_B} = \sqrt{\frac{E^2}{P^2} - \frac{1}{\ell^2}}\,r.
	\label{eq:killingr}
\ee
Similarly, using the Killing vector $k_T$ into Eq.~\eqref{eq:nullgeodesic} and integrating the resulting expression gives
\be
	r = \frac{P}{E\ell^2}\(T_B-T_A\).
	\label{eq:killingT}
\ee
Combining Eqs.~\eqref{eq:killingr} and~\eqref{eq:killingT} to get rid of the constants of motion, using Eq.~\eqref{eq:timesplit} and the identity
\be
	\frac{1}{R_A} - \frac{1}{R_B} = \int_{t_A}^{t_B} \frac{dt'}{R}H = \int_{R_A}^{R_B} \frac{dR'}{R^2},
\ee
results in
\be
	r_g^2 = \(\int_{t_A}^{t_B} \frac{dt'}{R} \sqrt{1 + \ell^2H^2}\)^2 - \(\int_{t_A}^{t_B} \frac{dt'}{R}\ell H\)^2.
    \label{eq:radius_grav}
\ee

Our Eq.~\eqref{eq:radius_grav} coincides with Eq.~(19) in Ref.~\cite{caldwell2001}. We express the integrals over $dt'$ in terms of the redshift of the object from the source,
\be
	1+z = \frac{R_B}{R_A},
\ee
where we assume that the source is located at redshift $z$, while the detector is located at redshift zero. Using the identity $dt/R = dR/R^2H = -dz/H$, we expand Eq.~\eqref{eq:radius_grav} around $\ell H \ll 1$, which is the low-energy limit in which the usual Friedmann equation is recovered. Keeping only terms up to order $\ell^2$, we obtain
\be
	r_g^2 \approx \(\int_0^z \frac{dz'}{H}\)^2 + \ell^2\(\int_0^z \frac{dz'}{H}\)\(\int_0^z dz'H\) - \ell^2 z^2.
    \label{eq:radius_grav_approx1}
\ee
The first integral expression corresponds to the distance traveled by the luminous signal on the brane,
\be
	r_\gamma = \int_{t_A}^{t_B}\,\frac{dt'}{R} = \int_0^z\,\frac{dz'}{H(z')}.
\ee
To approximate the second integral in Eq.~\eqref{eq:radius_grav_approx1}, we use the expression for the Hubble rate valid after matter-radiation equality, within a flat $\Lambda$CDM scenario,
\be
	H \simeq H_0[\Omega_m(1+z)^3 + \Omega_{\Lambda}]^{1/2},
    \label{eq:friedmann_matter}
\ee
where $\Omega_\Lambda = 1-\Omega_m$, with $\Omega_m$ and $\Omega_{\Lambda}$ the matter and dark energy density parameters respectively. Further assuming that the source lies within $z \ll 1$ gives, it can be expanded around $z=0$,
\bea
	\int_0^z dz'\frac{H_0}{H} &\approx& z - \frac{3 \Omega_m z^2}{4} - \[\frac{\Omega_m}{2} - \frac{9 \Omega_m^2}{8}\] z^3,\\
    \int_0^z dz' \frac{H}{H_0} &\approx& z + \frac{3 \Omega_m z^2}{4} + \[\frac{\Omega_m}{2} - \frac{3 \Omega_m^2}{8}\] z^3.
\eea
Finally, Eq.~\eqref{eq:radius_grav_approx1} is approximated at low redshift and small curvature radius $H\ell \ll 1$:
\be
	r_g^2 \approx r_\gamma^2 + \ell^2\frac{3 \Omega_m^2 z^4}{16},\quad\hbox{or}\quad \left|\frac{r_g - r_\gamma}{r_\gamma}\right| \approx \frac{3\ell^2 \Omega_m^2 z^4}{32r_\gamma^2}.
    \label{eq:radius_grav_approx2}
\ee
This expression differs from that obtained in Ref.~\cite{caldwell2001}, where the results are given for a single fluid component with equation of state $\omega = P/\rho$. Here instead, we specify the flat $\Lambda$CDM Hubble rate in Eq.~\eqref{eq:friedmann_matter}. The contribution to the time-delay given by the Shapiro delay~\cite{Shapiro1964} is discussed in the Appendix and it is found to be subdominant.

\section{Analysis method}
\label{sec:analysis}

The LIGO and Virgo collaborations have just announced that the NS-NS merger event with corresponding gravitational wave emission GW170817 occurred at a luminosity distance $r_g=40^{+8}_{-14}\,$Mpc~\cite{NS2017, NS20171}. Notice that at the very low redshifts under consideration ($z\ll0.1$), the luminosity distances and comoving distances approximately coincide. The EM signal GRB170817A measured by the \textit{Fermi} Gamma-ray Burst Monitor and the \textit{INTEGRAL} Anti-Coincidence Shield spectrometer arrived within a time-lag $\dtm = (1.734 \pm 0.054)\,$s from the detection of the associated GW signal~\cite{NS2017, NS20171}. The time-lag has been used to probe exotic physics such as modified gravity scenarios~\cite{lombriser2016a, lombriser2016b, bettoni, alonso, renk, creminelli2017, sakstein2017, ezquiaga2017, baker, NS2017, wei2017, green2017, arai2017, jana2017, gong2017, amendola2017, hou2017, crisostomivainshtein, dutta2017, langlois2017, peirone2017, kreisch2017, bartolo2017, babichev2017, dima2017, crisostomi2017, amendola2017new, linder2018, berti2018, cai2018, nersisyan2018,
kase2018, gong2018, granda2018, oost2018, tattersall2018, diez-tejedor2018, saltas2018, casalino2018, casalino2018new}.

Here, we use the measured time-lag to probe the size of extra dimensions, in particular the AdS$_5$ radius of curvature $\ell$. However, in performing this analysis it is of vital importance to keep in mind that \textit{most} of the time-lag is expected to be due to astrophysical processes involved in the merger of the two NSs. These processes, which are associated with the collapse of the hypermassive neutron star (HMNS)~\footnote{Often the collapse results in a rapidly rotating BH surrounded by a hot torus.}, in most cases result in the SGRB being emitted after the GW signal~\cite{NS2017}.  Therefore, it is hard to determine what proportion of the time-delay is due to the difference in the emission time of SGRB and GW signals, and to what extent that could result from differences in the propagation time of the two. It is also worth noting that the intergalactic medium dispersion is expected to have negligible impact on the EM propagation speed~\cite{NS2017}.

The above discussion makes it clear that quantifying the contribution of astrophysical processes to the measured time-delay is crucial to performing a correct analysis. We follow the approach of~\cite{NS2017}, where the size of astrophysical uncertainties associated with the measurement of the time-delay is conservatively quantified as $\simeq 10\,$s. This estimate of the astrophysical uncertainty is also the one used when deriving the constraints on the fractional GW-EM speed difference reported in~\cite{NS2017}. It is worth noticing that there exist more exotic models for binary NS merger events where the size of the time-lag due to astrophysical processes can be of the order of $100$-$1000\,$s~\cite{rezzolla2014,ciolfi2014} (although see~\cite{margalit2015} where the feasibility of the physical picture leading to time-lags of this order has been questioned). Conversely there are also models where the sign of the time-lag is reversed (i.e. the EM signal emission occurs before the GW signal)~\cite{tsang2011}. In most of these models, the larger time-lag is due to the binary NS merger producing a long-lived HMNS, or even a ``supramassive'' neutron star. However, the large amount of ejecta produced in the event that we are considering, as well as the absence of emissions powered by the NS spin-down, suggest that the HMNS produced collapsed within ${\cal O}(10^{-3}\,$s$)$. Hence, the exotic models described above are unlikely to correctly describe the binary NS merger associated to GW170817~\cite{margalit2017}.

In light of this discussion, we conclude that a safe and conservative choice for the size of the astrophysical uncertainties associated to the measured time-lag is $\sigma_{\rm astro} = 10\,$s; the same choice made in the joint LIGO/Virgo/\textit{Fermi}/\textit{INTEGRAL} analysis~\cite{NS2017}. Nevertheless, it is clear that a more precise characterization of the dynamics of binary NS mergers  (through ever-increasingly accurate magnetohydrodynamic simulations, e.g.~\cite{endrizzi}),  is necessary to separate astrophysical contributions to the EM-GW time-lag from those of exotic physics.

%

We constrain the size of the AdS$_5$ radius of curvature by performing a simple likelihood analysis on the available data. We sample the 4-dimensional parameter space spanned by the parameters $r_{\gamma}$, $\Omega_m$, $z$, and $\ell$, which we collectively refer to as $\boldsymbol{\theta}$. The available data consists of the time-lag measurement $\dtm$, which we refer to as $\boldsymbol{d}$. Our likelihood ${\cal L}$ then consists of the probability of observing the measured data, given the parameters $\boldsymbol{\theta}$: ${\cal L}(\boldsymbol{\theta}) = {\rm Pr}(\boldsymbol{d} \vert \boldsymbol{\theta})$.

We construct our likelihood ${\cal L}$ as an univariate Gaussian centered around $\delta t = \dtm$, i.e.:
\begin{eqnarray}
{\cal L}(\boldsymbol{\theta}) \!=\! \exp\!\! \left \{ \frac{\!-\![\delta t_{\rm th}(\boldsymbol{\theta}) \!-\! \delta t_{\rm meas}]^2}{2\sigma_{\rm tot}^2} \! \right \}.
\label{likelihood}
\end{eqnarray}
In Eq.~(\ref{likelihood}), $\delta t_{\rm th}$ denotes the theoretically expected value for the time-lag in the presence of extra dimensions, which is computed by identifying $r_g-r_{\gamma} \approx c\delta t$ in Eq.~(\ref{eq:radius_grav_approx2}):
\begin{eqnarray}
\delta t_{\rm th}(\boldsymbol{\theta}) \equiv \delta t_{\rm th}(\ell,\Omega_m,z,r_{\gamma}) = \frac{3\ell^2\Omega_m^2z^4}{32cr_{\gamma}} \, .
\label{deltattheo}
\end{eqnarray}
Finally, in Eq.~(\ref{likelihood}) we have indicated with $\sigma_{\rm tot}$ the \textit{total} uncertainty, comprising both the measurement uncertainty $\sigma_{\dtm} = 0.054\,$s, as well as the astrophysical uncertainty which we have quantified as $\sigma_{\rm astro} = 10\,$s following~\cite{NS2017}. Since the two uncertainties are completely independent (the first one is associated to the measurement process, whereas the second one is astrophysical in nature), we can combine them in quadrature to estimate the total uncertainty:
\begin{eqnarray}
\sigma_{\rm tot} = \sqrt{\sigma_{\dtm}^2+\sigma_{\rm astro}^2} \simeq 10\,{\rm s} \, ,
\end{eqnarray}
which, as expected given the difference in order of magnitude between the two, is entirely dominated by the size of the astrophysical uncertainty.

We impose a top-hat prior between 0 and 5 Mpc for $\ell$.~\footnote{We have numerically verified that the data, through the likelihood, cuts the distribution of $\ell$ well before 5 Mpc. Therefore the upper bound of the top-hat prior does not cut the distribution of $\ell$ where it is significantly non-zero.} We impose Gaussian priors on the remaining three parameters, conforming to their measured values. For $r_{\gamma}$ and $z$ we use the priors $(42.9\pm3.2)\, {\rm Mpc}$ and $z = 0.0080 \pm 0.0025$ respectively, as determined by the joint LIGO/Virgo/\textit{Fermi}/\textit{INTEGRAL} analysis in~\cite{NS20171} and consistent with the distance to the host galaxy of GW170817 (NGC4993)~\cite{NS20171}.~\footnote{Notice that the value of $r_{\gamma}$ provided in~\cite{NS20171} was evaluated combining the ratios of the Hubble flow velocity of NGC4993 to the two most widely used estimates of the Hubble constant.} We also use the prior inferred by the \textit{Planck} collaboration 2015 data release $\Omega_m = 0.315 \pm 0.013$~\footnote{Notice that, in principle, the large size of the extra dimension we will derive in Sec.~\ref{sec:results} might be expected to affect the interpretation of cosmological observations and correspondingly the parameters inferred by \textit{Planck}, including the adopted prior on $\Omega_m$. We defer further investigation of this issue to future work.}, coming from a combination of temperature and large-scale polarization data (\textit{Planck} TT+lowP)~\cite{planck2,planck3}. Notice that for $z$ we have used the symmetric error bar when imposing the prior, but have explicitly verified that using the non-symmetric error bar provided by the measurement has virtually no effect on our conclusions.

The posterior distribution of the parameters given the data is then constructed as the product of the likelihood and the priors. We sample the posterior distribution using Markov Chain Monte Carlo (MCMC) methods, by implementing the Metropolis-Hastings algorithm. We do so by using the cosmological MCMC sampler \texttt{Montepython}~\cite{montepython}, configured to act as a generic sampler. We use the generated chains to compute joint and marginalized posterior probability distributions of the four parameters and, in particular, of the curvature radius $\ell$. From the marginalized posterior distribution of $\ell$ we obtain the upper $68\%$ and $95\%$ confidence level (C.L.) upper limits on this quantity which we quote.

\section{Results}
\label{sec:results}

\begin{figure}[h!]
\includegraphics[width=1.0\linewidth]{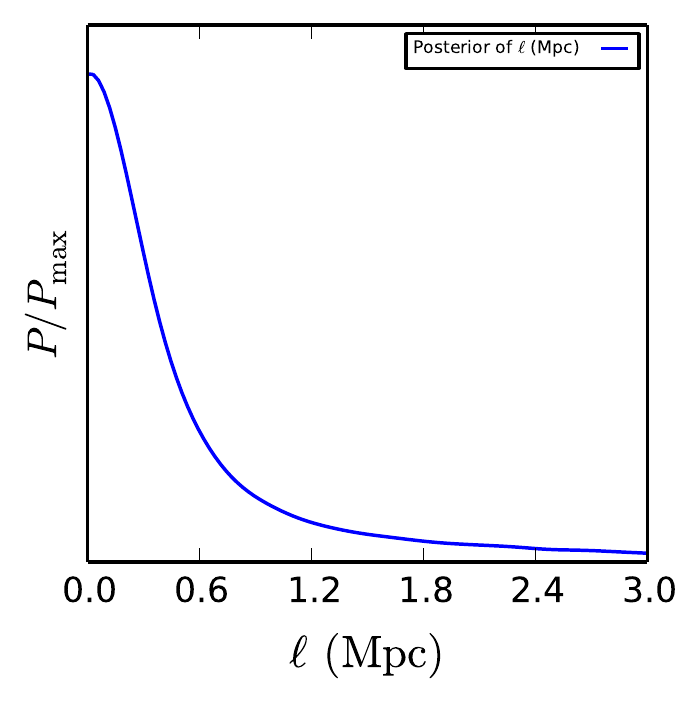}
\caption{Marginalized posterior distribution (normalized to its maximum value) of the AdS$_5$ radius of curvature $\ell$ in light of the time-lag between the GW170817 GW event and the corresponding EM counterpart GRB170817A.}
\label{fig:posterior}
\end{figure}

Here, we report the results of the likelihood analysis performed with the methodology described in the previous section. The posterior probability distribution we find for $\ell$ shown in Fig.~\ref{fig:posterior} is, as expected, sharply peaked at $0\,{\rm Mpc}$ and falls as $\ell$ increases. In particular, we find a $68\%$~C.L. upper limit of $\ell<0.535\,{\rm Mpc}$ and a $95\%$~C.L. upper limit of $\ell<1.997\,{\rm Mpc}$.

The upper bound of $\ell<0.535\,{\rm Mpc}$ at $68\%$~C.L. is a very poor constraint on the AdS$_5$ radius of curvature, since it is well known from experiments within the Solar System that Newtonian gravity works on the sub-millimiter scale~\cite{long2003, iorio2012, linares2014, tan2016}. However, the novelty of these results rests upon the fact that it is the \textit{very first time} that multi-messenger GW-EM astronomy is being used to probe the geometry of extra dimensions. Therefore, our results serve as an important proof of principle that multi-messenger astronomy can indeed be used to put constraints on the geometry of extra dimensions. We focused on the brane-world extra dimensional paradigm, however Refs.~\cite{gogberashvili2016, yu2017} put constraints on different models with extra-dimensions.

A natural question to ask is then: which aspects need the most improvement for the bound on the size of extra dimensions to be refined with future observations? Certainly a greater sample of multi-messenger GW-EM events beyond the one we have so far would help: assuming perfect control of systematics (which is clearly an idealized case), $N$ events would improve the uncertainty by $\approx \sqrt{N}$.

Another important possibility is that the parameters determining the physics of the time-lag (namely, $\Omega_m$, $z$, and $r_{\gamma}$) will eventually be measured to greater accuracy. For which of these parameters would an improved determination be especially useful for better constraining $\ell$? We can answer this question by examining the correlations between $\ell$ and the other 3 parameters in our likelihood analysis. Parameters which are more strongly correlated with $\ell$ will affect its marginalized posterior distribution more strongly.

\begin{figure} [t!]
\includegraphics[width=1.0\linewidth]{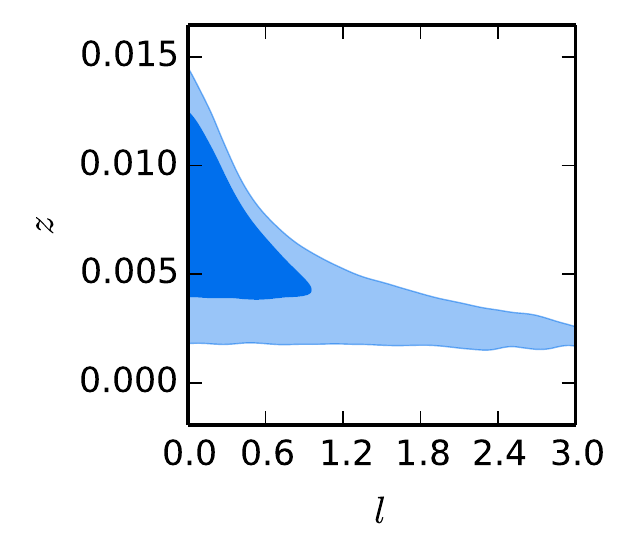}
\caption{Joint posterior distribution in the $\ell$-$z$ parameter space, with $\ell$ the AdS$_5$ radius of curvature and $z$ the redshift of the GW-EM event. The dark blue (light blue) regions correspond to $68\%$ ($95\%$)~C.L. contours respectively. From the figure it is clear that the two parameters are strongly negatively correlated, due to the fact that increasing/decreasing one and correspondingly decreasing/increasing the other can lead to the same value of the time-lag $\delta t$.}
\label{fig:jointposteriordistribution}
\end{figure}

We expect a strong inverse correlation between $\ell$ and $z$, since $\delta t$ depends on the combination $\ell^2z^4$ in Eq.~\eqref{deltattheo}. Therefore, it is possible to obtain the same $\delta t$ from various combinations of $\ell$ and $z$. In particular, if one of the two parameters is increased/decreased, the other will have to decrease/increase respectively in order to maintain $\delta t$ fixed. That is, the two parameters will be negatively correlated. In Fig.~(\ref{fig:jointposteriordistribution}), we plot the 2D joint posterior distribution in the $\ell$-$z$ parameter space. The likelihood analysis confirms our expectation that $\ell$ and $z$ are negatively correlated, which can be inferred by the orientation of the $\ell$-$z$ contour.

\begin{figure}
\includegraphics[width=1.0\linewidth]{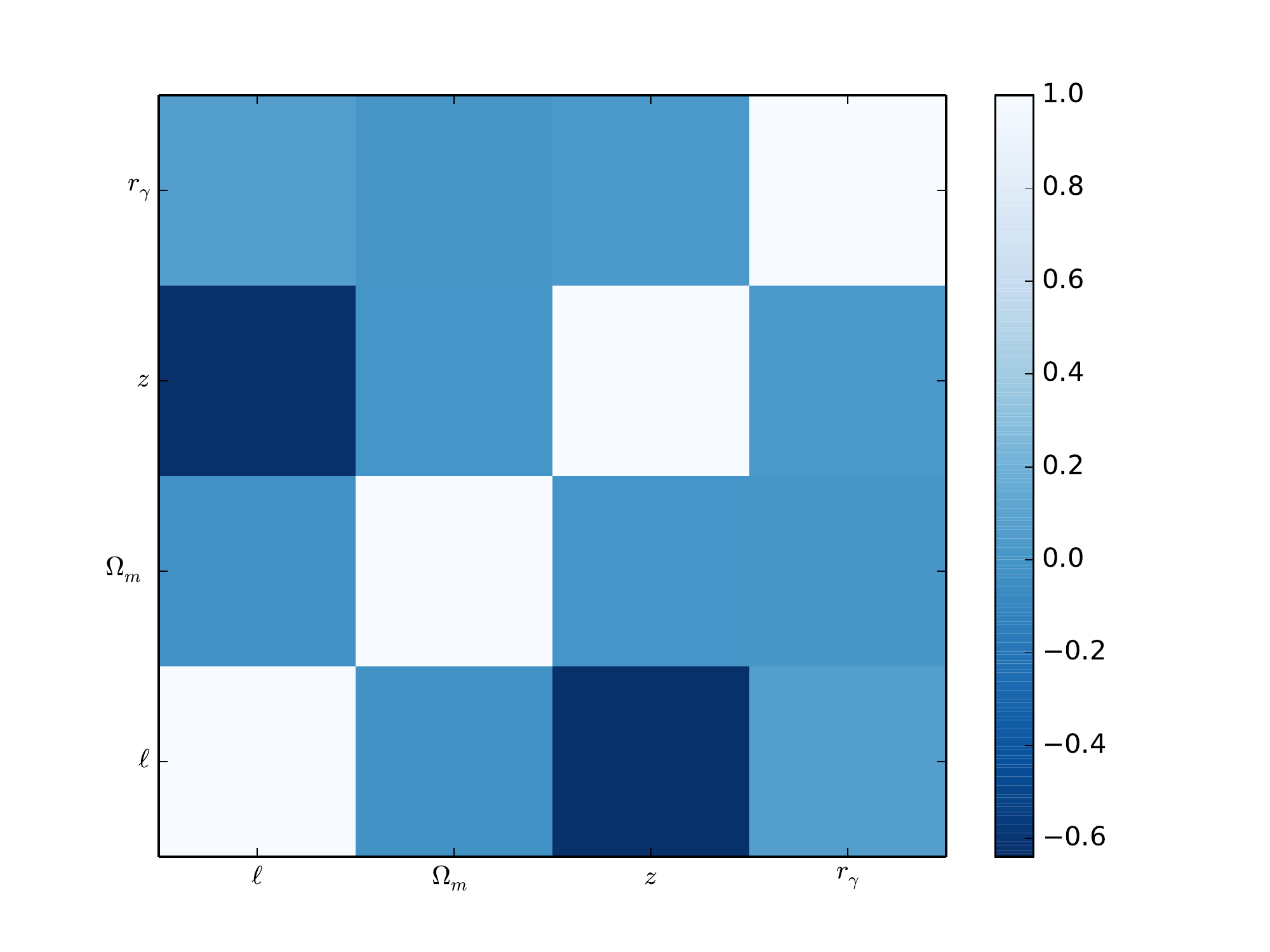}
\caption{Heatmap of the correlation matrix of the 4 parameters ($\ell$, $\Omega_m$, $z$, and $r_{\gamma}$) we are examining. We visually see that the strongest correlation is the negative correlation between $\ell$ and $z$, discussed in the text (see also the caption of Fig.~\ref{fig:jointposteriordistribution}).}
\label{fig:correlations}
\end{figure}

We also compute the correlation coefficients among each of the 4 parameters in $\boldsymbol{\theta}$. The correlation coefficient between two parameters $i$ and $j$, $R_{ij}$, is given by:
\begin{eqnarray}
R_{ij} \equiv \frac{C_{ij}}{\sqrt{C_{ii}C_{jj}}} \, ,
\end{eqnarray}
with $C$ being the covariance matrix of the cosmological parameters, estimated from the MCMC chains. In Fig.~(\ref{fig:correlations}), we plot a heatmap of the correlation matrix. The strongest correlation among the parameters is that between $\ell$ and $z$, with a magnitude of about $-0.6$. We conclude that an improved determination of the redshift of future multi-messenger GW-EM events will be especially useful for obtaining more stringent bounds on the physics of extra dimensions. The use of multiple detectors will prove extremely helpful in this direction.

The next-to-strongest correlation is that between $\ell$ and $\Omega_m$, which are also negatively correlated. Here, improvements in the determination of the matter energy density $\Omega_m$ will be possible thanks to measurements of the CMB temperature, polarization, and lensing anisotropy spectra from future ground-based CMB experiments such as Simons Observatory~\cite{simonsobservatory} and CMB-S4~\cite{s4}, in combination with BAO and clustering  measurements (matter power spectrum and/or shear power spectrum) from future galaxy redshift surveys and future weak lensing surveys such as DESI~\cite{desi}, LSST~\cite{lsst}, and Euclid~\cite{euclid}, as well as measurements of cross-correlations between CMB lensing convergence and galaxy clustering (which will improve the determination of $\Omega_m$ due to the improved determination of parameters which are mildly degenerate with it, such as $\sigma_8$, see e.g.~\cite{giusarma2018}). Therefore, improvements in the determination of cosmological parameters will also help to better constrain the physics of extra dimensions using multi-messenger GW-EM events.

\section{Conclusions}
\label{sec:conclusions}

The search for extra dimensions is motivated by theoretical speculations on the nature of our space-time and of gravity. It is possible that our four-dimensional universe might consist of the boundary to a ``bulk'' space where gravity is the only force which can propagate. In some models, GWs propagate through the bulk, taking a ``shortcut'' through the extra dimension, while EM signals travel on a null-geodesic confined to the brane. A consequence of such models is that a simultaneous event such as the emission of GWs and EM radiation from the merging of a compact object binary would be detected at Earth with a time-lag due to the different path of the two signals, despite their propagating with the same speed. The LIGO/Virgo collaboration measured the time-lag between the GW event GW170817~\cite{NS2017, NS20171} and the corresponding EM counterpart, the SGRB event GRB170817A detected by \textit{Fermi} and \textit{INTEGRAL}~\cite{NS2017}. Using this measurement, we investigated the five-dimensional brane-world scenario described in Ref.~\cite{caldwell2001} focusing on the flat $\Lambda$CDM model.

Performing a likelihood analysis which takes into account astrophysical uncertainties related to the emission of the SGRB associated to the GW event, as well as the measured time-lag, we have determined an upper bound of $\ell \lesssim 0.535\,$Mpc at $68\%$~C.L. on the radius of curvature of AdS$_5$. Although the bound obtained with this method is much weaker than the one obtained from measurements in the Solar System, the results aquired from the binary NS merger provide an independent test using data gathered from beyond our Solar System. Moreover, our results provide a proof-of-principle analysis applied to real data of the possibility of using multi-messenger astronomy to constrain the physics of extra dimensions. Our work also highlights the importance of robustly quantifying the contribution of astrophysical processes (such as the collapse time of the hypermassive neutron star generated by the merger) to the time-lag.

We remark that our analysis has been obtained within the $\Lambda$CDM model, assuming only one extra dimension exists. However, such an analysis can be extended to include more than one extra dimension, as well as metrics differing from AdS$_5$. The era of multi-messenger astronomy has just begun and it is exciting to notice how we can already use the available data to probe physics describing the very structure of our space-time, in the form of extra dimensions beyond those our senses are able to experience.

\begin{acknowledgments}
We thank Katherine Freese for useful discussions on the subject. SV thanks Thejs Brinckmann for assistance with the \texttt{Montepython} package, and Luciano Rezzolla for very useful correspondence. LV and SV acknowledge support by the Vetenskapsr\r{a}det (Swedish Research Council) through contract No. 638-2013-8993 and the Oskar Klein Centre for Cosmoparticle Physics. NB acknowledges funding from the European Research Council under the European Union's Seventh Framework Programme (FP7/2007-2013)/ERC Grant Agreement No. 617656 ``Theories and Models of the Dark Sector: Dark Matter, Dark Energy and Gravity.'' LV acknowledges support from the Central European Institute for Cosmology and Fundamental Physics (CEICO) in Prague, where part of this work was conducted.
\end{acknowledgments}

\section*{Note added} After our paper appeared on arXiv, the analysis of~\cite{pardo2018} was posted. In this work, constraints on the number of spacetime dimensions are placed by using the independent measurements of the luminosity distance to the GW source -- thanks to the fact that GWs are standard sirens -- and by measuring the redshift of the EM counterpart. We stress that the analysis in~\cite{pardo2018} is  different from ours and the results not applicable to our study focused on the brane-world model with an extra dimension of infinite size.

\appendix*

\section{Shapiro delay}
In this section we discuss the effect of the Shapiro delay \cite{Shapiro1964} on our analysis. The effects of the Shapiro delay on the GW170817 signal have been discussed in Refs.~\cite{wei2017, Wang2017, Boran2017, Shoemaker2017}. 

The line element for the AdS$_5$ spacetime in Eq.~\eqref{eq:RSmetric}, with the AdS$_5$ scale factor $f(R) = R^2/\ell^2$, can be rewritten as
\be
	ds^2 = \frac{R^2}{\ell^2}\left[ \eta_{\mu\nu}dx^\mu dx^\nu + \frac{\ell^4}{R^4}dR^2 \right],
	\label{eq:RSmetric1}
\ee
where $\eta_{\mu\nu}$ is the metric for a flat four-dimensional spacetime. It is possible to generalize the line element in Eq.~\eqref{eq:RSmetric1} by replacing $\eta_{\mu\nu}$ with any metric $g_{\mu\nu}$ that is a vacuum solution in General Relativity~\cite{Chamblin2000, Giannakis2001}. To quantify the importance of the Shapiro delay, we consider the Schwarzschild metric for $g_{\mu\nu}$, which describes the four-dimensional spacetime near a point mass $M$. The corresponding line element in Eq.~\eqref{eq:RSmetric1} reads~\cite{witten1998, Chamblin2000, Giannakis2001}
\be
	ds^2 = \frac{R^2}{\ell^2}\left[-U(r)dT^2 + U^{-1}(r)dr^2 + r^2d\Omega^2  + \frac{\ell^4}{R^4}dR^2 \right],
	\label{eq:RSmetric2}
\ee
where $U(r) = 1-2M/r$. This solution describes a mass line extending infinitely along the R-direction. Following the standard derivation for the Shapiro delay~\cite{Shapiro1964}, we look for a null-geodesic for which $ds^2 = 0$. The proper time on the bulk is
\be
	dT^2 \!=\! U^{-2}(r)dr^2 \!+\! U^{-1}(r)r^2d\Omega^2 \!+\! U^{-1}(r)\frac{\ell^4}{R^4}dR^2.
	\label{eq:RSmetric3}
\ee
The first two terms in Eq.~\eqref{eq:RSmetric3} reproduce the Shapiro delay experienced by the luminous signal propagating on the brane, 
\be 
	dt_{\text{brane}}^2 = \frac{dr^2}{(1-\frac{2M}{r})^2} +  \frac{ r^2 d\Omega^2}{(1-\frac{2M}{r})},
\ee 
while the extra term $U^{-1}(r)\frac{\ell^4}{R^4}dR^2$ describes the correction to the delay accrued by the gravitational wave signal. This latter correction is proportional to $(H \ell)^4\delta t^2$, thus it is of a higher order in $H\ell$ than the bulk effect we have discussed in Eq.~\eqref{eq:radius_grav_approx2}. We therefore conclude that the difference in Shapiro delay between the propagating light and gravitational waves is subdominant in our analysis.

\end{document}